12 March 2001    LBNL - 47066

# Historical roots of gauge invariance

J. D. Jackson *
*University of California and Lawrence Berkeley National Laboratory, Berkeley, CA 94720*

L. B. Okun †
*ITEP, 117218, Moscow, Russia*

**ABSTRACT**
Gauge invariance is the basis of the modern theory of electroweak and strong interactions (the so called Standard Model). The roots of gauge invariance go back to the year 1820 when electromagnetism was discovered and the first electrodynamic theory was proposed. Subsequent developments led to the discovery that different forms of the vector potential result in the same observable forces. The partial arbitrariness of the vector potential $\boldsymbol{A}$ brought forth various restrictions on it. $\boldsymbol{\nabla} \cdot \boldsymbol{A} = 0$ was proposed by J. C. Maxwell; $\partial_\mu A^\mu = 0$ was proposed L. V. Lorenz in the middle of 1860's. In most of the modern texts the latter condition is attributed to H. A. Lorentz, who half a century later was one of the key figures in the final formulation of classical electrodynamics. In 1926 a relativistic quantum-mechanical equation for charged spinless particles was formulated by E. Schrödinger, O. Klein, and V. Fock. The latter discovered that this equation is invariant with respect to multiplication of the wave function by a phase factor $exp(ie\chi/\hbar c)$ with the accompanying additions to the scalar potential of $-\partial\chi/c\partial t$ and to the vector potential of $\boldsymbol{\nabla}\chi$. In 1929 H. Weyl proclaimed this invariance as a general principle and called it Eichinvarianz in German and gauge invariance in English. The present era of non-abelian gauge theories started in 1954 with the paper by C. N. Yang and R. L. Mills.

## CONTENTS

___

* Electronic address: jdj@lbl.gov
† Electronic address: okun@heron.itep.ru

## I. INTRODUCTION

The principle of gauge invariance plays a key role in the Standard Model which describes electroweak and strong interactions of elementary particles. Its origins can be traced to Vladimir Fock (1926b) who extended the known freedom of choosing the electromagnetic potentials in classical electrodynamics to the quantum mechanics of charged particles interacting with electromagnetic fields. Equations (5) and (9) of Fock's paper are, in his notation,

$$\mathbf{A} = \mathbf{A}_1 + \nabla f$$
$$\phi = \phi_1 - \frac{1}{c}\frac{\partial f}{\partial t} \quad , \qquad \text{[Fock's (5)]}$$
$$p = p_1 - \frac{e}{c} f \quad ,$$

and
$$\psi = \psi_0 \, e^{\,2\pi i\, p/h} \quad . \qquad \text{[Fock's (9)]}$$

In present day notation we write

$$\mathbf{A} \to \mathbf{A}' = \mathbf{A} + \nabla \chi \quad , \tag{1a}$$
$$\Phi \to \Phi' = \Phi - \frac{1}{c}\frac{\partial \chi}{\partial t} \quad , \tag{1b}$$
$$\psi \to \psi' = \psi \, exp(ie\,\chi/\hbar c) \quad . \tag{1c}$$

Here $\mathbf{A}$ is the vector potential, $\Phi$ is the scalar potential, and $\chi$ is known as the gauge function. The Maxwell equations of classical electromagnetism for the electric and magnetic fields are invariant under the transformations (1a,b) of the potentials. What Fock discovered was that, for the quantum dynamics, that is, the form of the quantum equation, to remain unchanged by these transformations, the wave function is required to undergo the transformation (1c), whereby it is multiplied by a local (space-time dependent) phase. The concept was declared a general principle and "consecrated" by Hermann Weyl ( 1928, 1929a, 1929b). The invariance of a theory under combined transformations such as (1,a,b,c) is known as a gauge invariance or a gauge symmetry and is a touchstone in the creation of modern gauge theories.

The gauge symmetry of Quantum Electrodynamics (QED) is an abelian one, described by the U(1) group. The first attempt to apply a non-abelian gauge symmetry SU(2) x SU(1) to electromagnetic and weak interactions was made by Oscar Klein (1938). But this prophetic paper was forgotten by the physics community and never cited by the author himself.

The proliferation of gauge theories in the second half of the 20th century began with the 1954 paper on non-abelian gauge symmetries by Chen-Ning Yang and Robert L. Mills (1954). The creation of a non-abelian electroweak theory by Glashow, Salam, and Weinberg in the 1960s was an important step forward, as were the technical developments by 't Hooft and Veltman concerning dimensional regularization and renormalization. The discoveries at CERN of the heavy W and Z bosons in 1983 established the essential correctness of the electroweak theory. The very extensive and detailed measurements in high-energy electron-positron collisions at CERN and at SLAC, and in proton-antiproton collisions at Fermilab and in other experiments have brilliantly verified the electroweak theory and determined its parameters with precision.

In the 1970's a non-abelian gauge theory of strong interaction of quarks and gluons was created. One of its creators, Murray Gell-Mann, gave it the name Quantum Chromodynamics (QCD). QCD is based on the SU(3) group, as each quark of given "flavor" (u, d, s, c, b, t) exists in three varieties or different "colors" (red, yellow, blue). The quark colors are analogues of electric charge in electrodynamics. Eight colored gluons are analogues of the photon. Colored quarks and gluons are confined within numerous colorless hadrons.

QCD and Electroweak Theory form what is called today the Standard Model, which is the basis of all of physics except for gravity. All experimental attempts to falsify the Standard Model have failed up to now. But one of the cornerstones of the Standard Model still awaits its experimental test. The search for the so-called Higgs boson (or simply, higgs) or its equivalent is of profound importance in particle physics today. In the Standard Model, this electrically neutral, spinless particle is intimately connected with the mechanism by which quarks, leptons and W-, Z-bosons acquire their masses. The mass of the higgs itself is not restricted by the Standard Model, but general theoretical arguments imply that the physics will be different from expected if its mass is greater than 1 $TeV/c^2$. Indirect indications from LEP experimental data imply a much lower mass, perhaps 100 $GeV/c^2$, but so far there is no direct evidence for the higgs. Discovery and study of the higgs was a top priority for the aborted Superconducting Super Collider (SSC). Now it is a top priority for the Large Hadron Collider (LHC) under construction at CERN.

The key role of gauge invariance in modern physics makes it desirable to trace its historical roots, back to the beginning of 19th century. This is the main aim of our review. In section II, the central part of our article, we describe the history of classical electrodynamics with a special emphasis on the freedom of choice of potentials **A** and Φ expressed in equations (1a) and (1b).

It took almost a century to formulate this non-uniqueness of potentials that exists despite the uniqueness of the electromagnetic fields. The electrostatic potential resulting from a distribution of charges was intimately associated with the electrostatic potential energy of those charges and had only the trivial arbitrariness of the addition of a constant. The invention of Leyden jars and the development of voltaic piles led to study of the flow of electricity and in 1820 magnetism and electricity were brought together by Oersted's discovery of the influence of a nearby current flow on a magnetic needle (Jelved, Jackson, and Knudsen, 1998). Ampère and others rapidly explored the new phenomenon, generalized it to current-current interactions, and developed a mathematical description of the forces between closed circuits carrying steady currents (Ampère, 1827). In 1831 Faraday made the discovery that a time varying magnetic flux through a circuit induces current flow (Faraday, 1839). Electricity and magnetism were truly united.

The lack of uniqueness of the scalar and vector potentials arose initially in the desire of Ampère and others to reduce the description of the forces between actual current loops to differential expressions giving the element of force between infinitesimal current elements, one in each loop. Upon integration over the current flow in each loop, the total force would result. Ampère believed the element of force was central, that is, acting along the line joining the two elements of current, but others wrote down different expressions leading to the same integrated result. In the 1840's the work of Neumann (1847, 1849) and Weber (1878, 1848) led to competing differential expressions for the elemental force, Faraday's induction, and the energy between current elements, expressed in terms of different forms for the vector potential **A**. Over 20 years later Helmholtz (1870) ended the controversy by showing that Neumann's and Weber's forms for **A** were physically equivalent. Helmholtz's linear combination of the two forms with an arbitrary coefficient is the first example of what we now call a restricted class of different gauges for the vector potential.

The beginning of the last third of the 19th century saw Maxwell's masterly creation of the correct complete set of equations governing electromagnetism, unfortunately expressed in a way that many found difficult to understand (Maxwell, 1865). Immediately after, the Danish physicist Ludvig V. Lorenz, apparently independently of Maxwell, brilliantly developed the same basic equations and conclusions about the kinship of light and the electromagnetism of charges and currents (Lorenz, 1867b). From the point of view of gauge invariance, Lorenz's contributions are most significant. He introduced the so-called retarded scalar and vector potentials and showed that they satisfied the relation almost universally known as the Lorentz condition, though he preceded the Dutch physicist H. A. Lorentz by more that 25 years.

By the turn of the century, thanks to, among others, Clausius, Heaviside, Hertz, and Lorentz, who invented what we now call microscopic electromagnetism, with localized charges in motion forming currents, the formal structure of electromagnetic theory, the role of the potentials, the interaction with charged particles, the concept of gauge transformations, not yet known by that name, were in place. Lorentz's encyclopedia articles (Lorentz, 1904a, b) and his book (Lorentz, 1909) established him as an authority in classical electrodynamics, to the exclusion of earlier contributors such as Lorenz.

The start of the 20th century saw the beginning of the quantum, of special relativity, and of radioactive transformations. In the 1910s attention turned increasingly to atomic phenomena, with the confrontation between Bohr's early quantum theory and experiment. By the 1920s the inadequacies of the Bohr theory were apparent. In 1925-1926, Heisenberg, Schrödinger, Born, and others invented quantum mechanics. Inevitably, when the interaction of charged particles with time varying electromagnetic fields came to be considered in quantum mechanics, the issue of the arbitrariness of the potentials would arise. What was not anticipated was how the consequences of a change in the electromagnetic potentials on the quantum mechanical wave function became transformed into a general principle that defines what we now call quantum gauge fields.

In Section III we review the extension of the concept of gauge invariance in the early quantum era, the period from the end of the first World War to 1930, with emphasis on the annus mirabilis, 1926. As in Section II, we explore how and why priorities for certain concepts were taken from the originators and bestowed on others. We also retell the well-known story of the origin of the term "gauge transformation." In Section IV we discuss briefly the physical meaning of gauge invariance and describe the plethora of different gauges in sometime use today, but leave the detailed description of subsequent developments to others.

In writing this article we mainly relied on original articles and books, but in some instances we used secondary sources (historical reviews and mongraphs). Among the many sources on the history of electromagnetism in the 19th century we mention Whittaker (1951), Reiff and Sommerfeld (1902), Hunt (1991), Darrigol (2000), Buchwald (1985, 1989, 1994), and Rosenfeld (1957). Volume 1 of Whittaker (1951) surveys all of classical electricity and magnetism. Reiff and Sommerfeld (1902) provide an early review of some facets of the subject from Coulomb to Clausius. As his title implies, Hunt (1991) focuses on the British developments from Maxwell to 1900, in particular, the works of George F. FitzGerald, Oliver Lodge, Oliver Heaviside, and Joseph Larmor, as Maxwell's theory evolved into the differential equations for the fields that we know today. Darrigol (2000) covers the development of theoretical and experimental electromagnetism in the 19th century, with emphasis on the contrast between Britain and the Continent in the interpretations of Maxwell's theory. Buchwald (1985) describes in detail the transition in the last quarter of the 19th century from the macroscopic electromagnetic theory of Maxwell to the microscopic theory of Lorentz and others. Buchwald (1989) treats early theory and experiment in optics in the first part of the 19th century. Buchwald (1994) focuses on the experimental and theoretical work of Heinrich Hertz as he moved from Helmholtz′s pupil to independent authority with a different world view. Rosenfeld′s essay (Rosenfeld, 1957) focuses on the mathematical and philosophical development of electrodynamics from Weber to Hertz, with special emphasis on Lorenz and Maxwell. His comments on Lorenz′s "modern" outlook are very similar to ours. None of these works stress the development of the idea of gauge invariance.

The early history of *quantum gauge theories* as well as more recent developments have been extensively documented (Okun, 1986; Yang, 1986; Yang, 1987; O′Raifeartaigh, 1997; O′Raifeartaigh and Straumann, 2000).

Ludvig Valentin Lorenz of the classical era and Vladimir Aleksandrovich Fock emerge as physicists given less than their due by history. The many accomplishments of Lorenz in electromagnetism and optics are summarized by Kragh (1991, 1992) and more generally by Pihl (1939, 1972). Fock′s pioneering researches have been described recently, on the occasion of the 100th anniversary of his birth (Novozhilov and Novozhilov, 1999, 2000; Prokhorov, 2000).

The word "gauge" was not used in English for transformations such as (1,a,b,c) until 1929 (Weyl, 1929a). It is convenient, nevertheless, to use the modern terminology even when discussing the works of 19th century physicists. Similarly, we usually write equations in a consistent modern notation, using Gaussian units for electromagnetic quantities.

## II. CLASSICAL ERA

### A. Early history - Ampère, Neumann, Weber

On 21 July 1820 Oersted announced to the world his amazing discovery that magnetic needles were deflected if an electric current flowed in a circuit nearby, the first evidence that electricity and magnetism were related (Jelved, Jackson, and Knudsen, 1998). Within weeks of the news being spread, experimenters everywhere were exploring, extending, and making quantitative Oersted′s observations, nowhere more than in France. In the fall of 1820, Biot and Savart studied the force of a current-carrying long straight wire on magnetic poles and announced their famous law - that for a given current and pole strength, the force on a pole was perpendicular to the wire and to the radius vector, and fell off inversely as the perpendicular distance from the wire (Biot and Savart, 1820). On the basis of a calculation of Laplace for the straight wire and another experiment with a V-shaped wire, Biot abstracted the conclusion that the force on a pole exerted by an increment of the current of length *ds* was (a) proportional to the product of the pole strength, the current, the length of the segment, the square of the inverse distance *r* between the segment and the pole, and to the sine of the angle between the direction of the segment and the line joining the segment to the pole, and (b) directed perpendicular to the plane containing those lines. (Biot, 1824). We recognize this as the standard expression for an increment of magnetic field *d**B*** times a pole strength - see for example Eq.(5.4), p 175 of Jackson (1998).

At the same time Ampère, in a brilliant series of demonstrations before the French Academy, showed, among other things, that small solenoids carrying current behaved in the Earth′s magnetic field as did bar magnets, and began his extensive quantitative observations of the forces between closed circuits carrying steady currents. These continued over several years; the papers were collected in a memoir in 1826 (Ampère, 1827).

The different forms for the vector potential in classical electromagnetism arose from the competing versions of the elemental force between current elements abstracted from Ampère's extensive observations. These different versions arise because of the possibility of adding perfect differentials to the elemental force, expressions that integrate to zero around closed circuits or circuits extending to infinity. Consider the two closed circuits $C$ and $C'$ carrying currents $I$ and $I'$, respectively, as shown in Fig. 1. Ampère believed that the force increment $d\mathbf{F}$ between differential directed current segments $I\,d\mathbf{s}$ and $I'd\mathbf{s}'$ was a central force, that is, directed along the line between the segments. He wrote his elemental force law in compact form (Ampère, 1827, p. 302),

$$dF = 4k\frac{II'}{\sqrt{r}}\frac{\partial^2 \sqrt{r}}{\partial s\, \partial s'}\,ds\,ds' = k\frac{II'}{r^2}\left[2r\frac{\partial^2 r}{\partial s\, \partial s'} - \frac{\partial r}{\partial s}\frac{\partial r}{\partial s'}\right]ds\,ds' \quad, \tag{2}$$

where the constant $k = 1/c^2$ in Gaussian units. The distance $r$ is the magnitude of $\mathbf{r} = \mathbf{x} - \mathbf{x}'$, where $\mathbf{x}$ and $\mathbf{x}'$ are the coordinates of $d\mathbf{s} = \mathbf{n}\,ds$ and $d\mathbf{s}' = \mathbf{n}'\,ds'$. In what follows we also use the unit vector $\hat{\mathbf{r}} = \mathbf{r}/r$. In vector notation and Gaussian units, Ampère's force reads

$$d\mathbf{F} = \frac{II'}{c^2}\frac{\hat{\mathbf{r}}}{r^2}\left[3\,\hat{\mathbf{r}}\cdot\mathbf{n}\,\hat{\mathbf{r}}\cdot\mathbf{n}' - 2\,\mathbf{n}\cdot\mathbf{n}'\right]ds\,ds' \quad. \tag{3}$$

It is interesting to note that Ampère has the equivalent of this expression at the bottom of p. 253 in (Ampère, 1827) in terms of the cosines defined by the scalar products. He preferred, however, to suppress the cosines and express his result in terms of the derivatives of $r$ with respect to $ds$ and $ds'$, as in (2).

The first observation to make is that the abstracted increment of force $d\mathbf{F}$ has no physical meaning because it violates the continuity of charge and current. Currents cannot suddenly materialize, flow along the elements $d\mathbf{s}$ and $d\mathbf{s}'$ and then disappear again. The expression is only an intermediate mathematical construct, perhaps useful, perhaps not, in finding actual forces between real circuits. The second observation is that the form widely used at present (see Eq.(5.8), p. 177 of Jackson (1998) for the integrated expression),

$$d\mathbf{F} = \frac{II'}{c^2 r^2}\mathbf{n}\times(\mathbf{n}'\times\hat{\mathbf{r}})\,ds\,ds' = \frac{II'}{c^2 r^2}\left[\mathbf{n}'(\hat{\mathbf{r}}\cdot\mathbf{n}) - \hat{\mathbf{r}}(\mathbf{n}\cdot\mathbf{n}')\right]ds\,ds' \quad, \tag{4}$$

was first written down independently in 1845 by Neumann (Eq.(2), p. 64 of Neumann, 1847) and Grassmann (1845)*. Although not how these authors arrived at it, one way to

———————————————

* A cogent discussion of Ampère's work, his running dispute with Biot, and also Grassmann's criticisms and alternative expression for the force, is given in the little book by Tricker (1965). Tricker also gives translations of portions of the papers by Oersted, Ampère, Biot and Savart, and Grassmann.

———————————————

understand its form is to recall that a charge $q'$ in nonrelativistic motion with velocity $\mathbf{v}'$ (think of a quasi-free electron moving through the stationary positive ions in a conductor) generates a magnetic field ($\boldsymbol{B'} \propto q'\boldsymbol{v'} \times \boldsymbol{r}/r^3$). Through the Lorentz force law $\boldsymbol{F} = q(\boldsymbol{E}' + \boldsymbol{v} \times \boldsymbol{B'}/c)$, this field produces a force on a similar charge $q$ moving with velocity $\boldsymbol{v}$ in a second conductor. Now replace $q\boldsymbol{v}$ and $q'\boldsymbol{v'}$ with $I\boldsymbol{n}ds$ and $I'\boldsymbol{n'}ds'$. With its non-central contribution, it does not agree with Ampère's, but the differences vanish for the total force between two closed circuits, the only meaningful thing. In fact, the first term in (4) contains a perfect differential, $\boldsymbol{n}\cdot\hat{\boldsymbol{r}}/r^2\, ds = -d\boldsymbol{s}\cdot\boldsymbol{\nabla}(1/r)$, which gives a zero contribution when integrated over the closed path $C$ in Fig. 1. If we ignore this part of $d\boldsymbol{F}$, the residue appears as a central force (!) between elements, although not the same as Ampère's central force.

Faraday's discovery in 1831 of electromagnetic induction - relative motion of a magnet near a closed circuit induces a momentary flow of current - exposed the direct link between electric and magnetic fields (Faraday, 1839). The experimental basis of quasi-static electromagnetism was now established, although the differential forms of the basic laws were incomplete and Maxwell's completion of the description with the displacement current was still 34 years in the future. Research tended to continue on the behaviour of current-carrying circuits interacting with magnets or other circuits. While workers spoke of induced currents, use of Ohm's law made it clear that they had in mind induced electric fields along the circuit elements.

Franz E. Neumann in 1845 and 1847 analyzed the process of electromagnetic induction in one circuit from the relative motion of nearby magnets and other circuits (Neumann, 1847, 1849). He is credited by later writers as having invented the vector potential, but his formulas are always for the induced current or its integral and so are products of quantities among which one can sense the vector potential or its time derivative lurking, without explicit display. In the latter parts of his papers, he adopts a different tack. As mentioned above, he expresses the elemental force between current elements in what amounts to (4). He then omits the perfect differential to arrive at an expression for the elemental force $d\boldsymbol{F}$ (the second term in (4)) that is the negative gradient with respect to $\boldsymbol{r}$ of a magnetic potential energy $dP$. From (4) we see that $dP$ and its double integral $P$ (over the circuits in Fig. 1) are

$$dP = -\frac{II'}{c^2}\frac{\boldsymbol{n}\cdot\boldsymbol{n'}}{r}\,ds\,ds' \quad ; \quad P = -\frac{II'}{c^2}\oiint_{C\,C'}\frac{\boldsymbol{n}\cdot\boldsymbol{n'}}{r}\,ds\,ds' \quad , \tag{5}$$

The double integral in (5) is the definition of the mutual inductance of the circuits $C$ and $C'$. The force on circuit $C$ is now the negative gradient of $P$ with respect to a suitable coordinate defining the position of $C$, with both circuits kept fixed in orientation. Neumann's $P$ is the negative of the magnetic interaction energy $W$, defined nowadays as

$$W = \frac{I}{c}\oint_C \boldsymbol{n}\cdot\boldsymbol{A}'\,ds \quad , \quad \text{with} \quad \boldsymbol{A}'(\boldsymbol{x}) = \frac{I'}{c}\oint_{C'} \frac{\boldsymbol{n'}}{r}\,ds' \quad , \tag{6}$$

where $\boldsymbol{A}'$ is the vector potential of the current $I'$ flowing in circuit $C'$. For a general current density $\boldsymbol{J}(\boldsymbol{x'}, t)$, this form of the vector potential is

$$\mathbf{A}_N(\mathbf{x}, t) = \frac{1}{c} \int d^3x' \frac{1}{r} \mathbf{J}(\mathbf{x}', t) \quad . \tag{7}$$

We have attached a subscript *N* to *A* here to associate it with Neumann's work, as did subsequent investigators, even though he never explicitly displayed (6) or (7).

Independently and at roughly the same time as Neumann, in 1846 Wilhelm Weber presented a theory of electromagnetic induction, considering both relative motion and time-varying currents as sources of the electromotive force in the secondary circuit. To this end, he introduced a central force law between two charges *e* and *e′* in motion, consistent with Ampère's law for current-carrying circuits (Weber, 1878, 1848). Weber adopted the hypothesis that current flow in a wire consists of equal numbers of charges of both signs moving at the same speed, but in opposite directions, rather than the general view at the time that currents were caused by the flow of two electrical fluids. He thus needed a basic force law between charges to calculate forces between circuits. Parenthetically we note that this hypothesis, together with the convention that the current flow was measured in terms of the flow of only one sign of charge, led to the appearance of factors of two and four in peculiar places, causing confusion to the unwary. We write everything with modern conventions. Weber's central force law, admittedly ad hoc and incorrect as a force between charges in motion, is (Weber, 1878, p. 229)

$$F = \frac{ee'}{r^2} + \frac{ee'}{c^2}\left[\frac{1}{r}\frac{d^2r}{dt^2} - \frac{1}{2r^2}\left(\frac{dr}{dt}\right)^2\right] \quad . \tag{8}$$

The first term is just Coulomb's law. The ingredients of the second part can be expressed explicitly as

$$\frac{dr}{dt} = \hat{\mathbf{r}}\cdot(\mathbf{v}-\mathbf{v}') \quad ; \quad \frac{d^2r}{dt^2} = \hat{\mathbf{r}}\cdot(\mathbf{a}-\mathbf{a}') + \frac{1}{r}\left[(\mathbf{v}-\mathbf{v}')^2 - (\hat{\mathbf{r}}\cdot(\mathbf{v}-\mathbf{v}'))^2\right] \quad , \tag{9}$$

where **v** and **a** (**v′** and **a′**) are the velocity and acceleration of the charge *e* (*e′*). If we add up the forces between the charges ±*e* with velocities ±**v** in the one current and those ±*e′* with velocities ±**v′** in the other, and identify 2*e***v** = *I***n**ds and 2*e′***v′** = *I′***n′**ds′, we obtain Ampère's expression (3).
If instead of the force between current elements, we consider the force on a charge at rest at **x**, the position of **n**ds, due to the current element *I′***n′**ds′, we find from Weber's force law,

$$d\mathbf{F} = -\frac{e}{c^2 r}\hat{\mathbf{r}}\hat{\mathbf{r}}\cdot\mathbf{n}'\frac{dI'}{dt}ds' \quad , \tag{10}$$

where *dI′/dt* arises from the presence of the acceleration **a′**. Weber's analysis was more complicated than that just described because he treated relative motion and time variation of the inducing current simultaneously, but for circuits with no relative motion Weber (Weber, 1848, p. 239) writes the induced electromotive force in the circuit, *dE = d***F**·**n**/e, in this form. Weber wrote only the component of the induced force or emf along the element **n**ds, but if we identify the induced electric field as *E* = - (1/c) ∂(d**A**)/∂t, with *d***A** the elemental vector potential, we find from (10) *d***A** and its integral **A** over the inducing circuit *C′* to be

$$d\mathbf{A} = \frac{I'}{cr}\hat{\mathbf{r}}\,\hat{\mathbf{r}}\cdot\mathbf{n}'\,ds'\quad;\qquad \mathbf{A} = \frac{I'}{c}\oint_C \frac{\hat{\mathbf{r}}\,\hat{\mathbf{r}}\cdot\mathbf{n}'}{r}\,ds'\quad. \tag{11}$$

The generalization of this form of the vector potential for a current density $\mathbf{J}(\mathbf{x}', t)$ is

$$\mathbf{A}_W(\mathbf{x}, t) = \frac{1}{c}\int d^3x'\,\frac{1}{r}\hat{\mathbf{r}}\,\hat{\mathbf{r}}\cdot\mathbf{J}(\mathbf{x}', t)\quad. \tag{12}$$

As with Neumann, we attach a subscript $W$ for Weber to this form of the vector potential even though he did not write (11) or (12) explicitly.

### B. Vector potentials - Kirchhoff and Helmholtz

Gustav Kirchhoff was the first to write explicitly (in component form) the vector potential (12); he also wrote the components of the induced current density as the conductivity times the negative sum of the gradient of the scalar potential and the time derivative of the vector potential (Kirchhoff, 1857, p. 530). He attributed the second term in the sum to Weber; the expression (12) became known as the Kirchhoff-Weber form of the vector potential. Kirchhoff applied his formalism to analyze the telegraph and calculate inductances.

We note in passing that Kirchhoff showed (contrary to what is implied by Rosenfeld, 1957) that the Weber form of $\mathbf{A}$ and the associated scalar potential $\Phi$ satisfy the relation (in modern notation), $\nabla\cdot\mathbf{A} = \partial\Phi/c\partial t$, the first published relation between potentials in what we now know as a particular gauge (Kirchhoff, 1857, p.532-533).

In an impressive, if repetitive, series of papers, Hermann von Helmholtz (1870, 1872, 1873, 1874) criticized and clarified the earlier work of Neumann, Weber, and others. He criticized Weber's force equation for leading to unphysical behavior of charged bodies in some circumstances, but recognized that Weber's form of the magnetic energy had validity. Helmholtz compared the Neumann and Weber forms of the magnetic energy between current elements, $dW = pII'\,ds\,ds'/c^2 r$, with $p(Neumann) = \mathbf{n}\cdot\mathbf{n}'$ and $p(Weber) = \mathbf{n}\cdot\hat{\mathbf{r}}\,\mathbf{n}'\cdot\hat{\mathbf{r}}$, and noted that they differ by a multiple of the perfect differential $ds\,ds'\,(\partial^2 r/\partial s\,\partial s') = ds\,ds'(\mathbf{n}\cdot\hat{\mathbf{r}}\,\mathbf{n}'\cdot\hat{\mathbf{r}} - \mathbf{n}\cdot\mathbf{n}')/r$. Thus either form leads to the same potential energy and force for closed circuits. Helmholtz then generalized the expressions of Weber and Neumann for the magnetic energy between current elements by writing a linear combination (Helmholtz, 1870, equation (1.), p. 76, but in modern notation),

$$dW = \frac{II'}{2\,c^2\,r}\left[\,(1+\alpha)\,\mathbf{n}\cdot\mathbf{n}' + (1-\alpha)\,\mathbf{n}\cdot\hat{\mathbf{r}}\,\mathbf{n}'\cdot\hat{\mathbf{r}}\,\right]ds\,ds'\quad. \tag{13}$$

Obviously, this linear combination differs from either Weber's or Neumann's expressions by a multiple of the above perfect differential, and so is consistent with Ampère's observations. The equivalent linear combination of the vector potentials (7) and (12) is (*ibid.,* equation (1[a].), p. 76, in compressed notation)

$$\mathbf{A}_\alpha = \frac{1}{2}(1 + \alpha)\, \mathbf{A}_N + \frac{1}{2}(1 - \alpha)\, \mathbf{A}_W \quad . \tag{14}$$

$\alpha = 1$ gives the Neumann form; $\alpha = -1$ gives Weber. Helmholtz's generalization exhibits a one-parameter class of potentials that is equivalent to a family of vector potentials of different gauges in Maxwell's electrodynamics. In fact, in equation (1d.) on p. 77, he writes the connection between his generalization and the Neumann form (7) as (in modern notation),

$$\mathbf{A}_\alpha = \mathbf{A}_N + \frac{(1 - \alpha)}{2} \nabla \Psi \quad , \quad \text{where} \quad \Psi = -\frac{1}{c} \int \hat{\mathbf{r}} \cdot \mathbf{J}(\mathbf{x}', t)\, d^3x' \quad . \tag{15}$$

Helmholtz goes on to show that $\Psi$ satisfies $\nabla^2 \Psi = 2\, \partial\Phi/c\, \partial t$, where $\Phi$ is the instantaneous electrostatic potential, and that $\Phi(\mathbf{x}, t)$ and his vector potential $\mathbf{A}_\alpha(\mathbf{x}, t)$ are related by (*ibid.*, equation ($3^a$.), p. 80, in modern notation)

$$\nabla \cdot \mathbf{A}_\alpha = -\alpha \frac{\partial \Phi}{c\, \partial t} \quad . \tag{16}$$

This relation contains the connection found in 1857 by Kirchhoff (for $\alpha = -1$) and formally the condition found in 1867 by Lorenz -see below - but Helmholtz's relation connects only the quasi-static potentials, while Lorenz's relation holds for the fully retarded potentials. Helmholtz is close to establishing the gauge invariance of electromagnetism, but treats only a restricted class of gauges and lacks the transformation of the scalar as well as the vector potential.

Helmholtz remarks rather imprecisely that the choice of $\alpha = 0$ leads to Maxwell's theory. The resulting vector potential,

$$\mathbf{A}_M(\mathbf{x},t) = \frac{1}{2c} \int \left( \frac{\mathbf{J}(\mathbf{x}', t)}{r} + \frac{\hat{\mathbf{r}}\hat{\mathbf{r}} \cdot \mathbf{J}(\mathbf{x}', t)}{r} \right) d^3x' \quad , \tag{17}$$

can be identified with Maxwell only because, as (16) shows, it is the quasi-static vector potential found from the transverse current for $\nabla \cdot \mathbf{A} = 0$, Maxwell's preferred choice for $\mathbf{A}$. Maxwell never wrote down (17). It is relevant for finding an approximate Lagrangian for the interaction of charged particles, correct to order $1/c^2$ - see Section II.D.

We see in the early history the attempts to extend Ampére's conclusions on the forces between current-carrying circuits to a comprehensive description of the interaction of currents largely within the framework of potential energy, in analogy with electrostatics. Competing descriptions stemmed from the arbitrariness associated with the postulated elemental interactions between current elements, an arbitrariness that vanished upon integration over closed circuits. These differences led to different but equivalent forms for the vector potential. The focus was on steady-state current flow or quasi-static behavior. Meanwhile, others were addressing the propagation of light and its possible connection with electricity, electric currents, and magnetism. That electricity was due to discrete charges and electric currents to discrete charges in motion was a minority view, with Weber a notable advocate. Gradually, those ideas gained credence and charged particle dynamics came under

4study. Our story now turns to these developments and how the concept of different gauges was elaborated, and by whose hands.

## C. Electrodynamics by Maxwell, Lorenz, and Hertz

The vector potential played an important role in Maxwell's emerging formulation of electromagnetic theory (Bork, 1967; see also Everitt, 1975). He developed an analytic description of Faraday's intuitive idea that a conducting circuit in a magnetic field was in an "electro-tonic state," ready to respond with current flow if the magnetic flux linking it changed in time (Maxwell, 1856). He introduced a vector, "electro-tonic intensity," with vanishing divergence, whose curl is the magnetic field **B** or equivalently whose line integral around the circuit is related by Stokes's theorem to the magnetic flux through the loop. Including both time-varying magnetic fields and motion of the circuit through an imhomogeneous field, Maxwell expressed the electromotive force (in our language) as $c\mathbf{E} = - d\mathbf{A}/dt = -\partial \mathbf{A}/\partial t + (\mathbf{v}\cdot\mathbf{\nabla})\mathbf{A}$ . Maxwell contrasts his mathematical treatment (which he does "not think [that it] contains even the shadow of a true physical theory") with that of Weber, which he calls "a professedly physical theory of electro-dynamics, which is so elegant, so mathematical, and so entirely different from anything in this paper,...." (Maxwell, 1856, *Sc. P.*, Vol. 1, p.207-208).

Interestingly, in the introduction to (Maxwell, 1865), while praising Weber and Carl Neumann, he distances himself from them in avoiding charged particles as sources, velocity-dependent interactions, and action-at-a-distance, preferring the mechanism of excited bodies and the propagation of effects through the ether. Specifically he states (Maxwell, 1865, *Sc. P.* Vol. 1, p. 528):

> "We therefore have some reason to believe, from the phenomena of light and heat, that there is an aethereal medium filling space and permeating bodies, capable of being set in motion and of transmitting that motion from one point to another, and of transmitting that motion to gross matter so as to heat it and affect it in various ways."

In this paper he again asserts his approach to the vector potential, now called "electromagnetic momentum," with its line integral around a circuit called the total electromagnetic momentum of the circuit. The central role of the vector potential in Maxwell's thinking is evidenced in the table (*op. cit.,* p. 561) where in his list of the 20 variable quantities in his equations *(F, G, H)* , the components of "Electromagnetic Momentum," top the list. A similar list in his treatise (Maxwell, 1873, Vol. 2, Art. 618, p.236) has the electromagnetic momentum second, after the coordinates of a point. In (Maxwell, 1865, *Sc. P.*, Vol. 1, p. 564), he explains the use of the term "electomagnetic momentum" as a result of the analogy of the mechanical $\mathbf{F} = d\mathbf{p}/dt$ and the electromagnetic $c\mathbf{E} = - d\mathbf{A}/dt$, but cautions that it is "to be considered as illustrative, not as explanatory." * A curiosity is Maxwell's use in his treatise (Maxwell, 1873) of at least three different expressions for the same

---

*The aptness of the term electomagnetic momentum goes beyond Maxwell's analogy; in the Hamiltonian dynamics of a charged particle the canonical momentum is $\mathbf{P} = \mathbf{p} + e\mathbf{A}/c$ .

---

quantity - vector potential (Sects. 405, 590, 617), electrokinetic momentum (Sects. 579, 590), electromagnetic momentum (Sects. 604, 618 ).

Helmholtz's identification of (17) with Maxwell is because Maxwell preferred $\mathbf{\nabla}\cdot\mathbf{A} = 0$ when using any vector potential (Maxwell, 1865, Sect. 98, p. 581; Maxwell, 1873, 1st ed., Sects. 616, 617, p. 235-236; 3rd ed., p. 256). In (Maxwell, 1873) he writes the vector potential **A′** in the Neumann form (7), but with the "total current," conduction **J** plus displacement $\partial \mathbf{D}/c\partial t$, instead of $\mathbf{J}$ alone. [We transcribe his notation into present day notation where appropriate.] He then writes what is now called the gauge transformation equation $\mathbf{A} = \mathbf{A'} - \nabla \chi$ (Maxwell, 1873, equation (7), Sect. 616, p. 235, 1st ed., p.256, 3rd ed.) and observes:

> "The quantity $\chi$ disappears from the equations (A) [$\mathbf{B} = \nabla \times \mathbf{A}$] and it is not related to any physical phenomenon."

He goes on to say that he will set $\chi = 0$, remove the prime from $\mathbf{A'}$ and have it as the true value of the vector potential. The virtue to Maxwell of his $\mathbf{A}$ is that

> "it is the vector-potential of the electric current, standing in the same relation to the electric current that the scalar potential stands to the matter of which it is the potential."

Maxwell's statement $\mathbf{A} = \mathbf{A'} - \nabla \chi$ and the invariance of the fields under this (gauge) transformation is one of the earliest explicit statements, more general than Helmholtz's, but he misses stating the accompanying transformation of the scalar potential because of his use of the "total current" as the source of the vector potential. In the quasi-static limit, the elimination of the displacement current in vacuum in favor of the potentials and their sources leads to (17), the form Helmholtz identified with Maxwell.

The Danish physicist Ludvig Valentin Lorenz is perhaps best known for his pairing with the more famous Dutch physicist Hendrik Antoon Lorentz in the Lorenz-Lorentz relation between index of refraction and density. In fact he was a pioneer in the theory of light and in electrodynamics, contemporaneous with Maxwell. In 1862 he developed a mathematical theory of light, using the basic known facts (transversality of vibrations, Fresnel's laws), but avoiding the (unnecessary, to him) physical modeling of a mechanistic ether* with bizarre

---

* Notable in this regard, but somewhat peripheral to our history of gauge invariance, was James MacCullagh's early development of a phenomenological theory of light as disturbances propagating in a novel form of the elastic ether, with the potential energy depending not on compression and distortion but only on *local rotation* of the medium in order to make the light vibrations purely transverse (MacCullagh, 1839; Whittaker, 1951, p. 141-4; Buchwald, 1985, Appendix 2). MacCullagh's equations correspond (when interpreted properly) to Maxwell's equations for free fields in anisotropic media. We thank John P. Ralston for making available his unpublished manuscript on MacCullagh's work.

---

properties in favor of a purely phenomenological model (Lorenz, 1863). Indeed, in a Danish publication (Lorenz, 1867a) he took a very modern sounding position on the luminiferous ether, saying,

> "The assumption of an ether would be unreasonable because it is a new non-substantial medium which has been thought of only because light was conceived in the same manner as sound and hence had to be a medium of exceedingly large elasticity and small density to explain the large velocity of light. ..... It is most unscientific to invent a new substance when its existence is not revealed in a much more definite way." (translation taken from Kragh, 1991, p. 4690).

That same year, two years after Maxwell (1865) but evidently independently, he published a paper entitled "On the Identity of the Vibrations of Light with Electric Currents," (Lorenz, 1867b). On p. 287, addressing the issue of the disparities between the nature of electricity (two fluids), light (vibrations of the ether), and heat (motion of molecules) half a

13century after Oersted's discoveries, he laments the absence of a unity of forces. He continues,

> "Hence it would probably be best to admit that in the present state of science we can form no conception of the physical reason of forces and of their working in the interior of bodies; and therefore (at present, at all events) we must choose another way, free from all physical hypotheses, in order, if possible, to develope (sic) theory step by step in such a manner that the further progress of a future time will not nullify the results obtained."

Avoiding the distasteful ether, Lorenz follows Kirchhoff in attributing a conductivity to material media, and also a negligibly small but not zero conductivity for "empty" space. He thus deals with current densities rather than electric fields, which he defines according to Ohm's law ($J = \sigma E$); many of his equations are the customary ones when divided by the conductivity $\sigma$. After stating Kirchhoff's version of the static potentials, in which the vector potential is Weber's form (12), he observes that retardation is necessary to account for the finite speed of propagation of light and, he supposes, electromagnetic disturbances in general. He generalizes the static scalar and vector potentials to the familiar expressions, often attributed to Lorentz, by introduction of $\bar{\Omega}$ ( Lorenz, 1867b, p.289, [Phil. Mag.]) as his scalar potential and $\alpha, \beta, \gamma$ ( *ibid.*, p. 291) as the components of his vector potential. In modern notation these are

$$\Phi(\mathbf{x}, t) = \int \frac{\rho(\mathbf{x}', t - r/c)}{r} d^3x' \; ; \qquad \mathbf{A}(\mathbf{x},t) = \frac{1}{c} \int \frac{\mathbf{J}(\mathbf{x}', t - r/c)}{r} d^3x' \; , \qquad (18)$$

the latter being the retarded form of the Neumann version (7). After showing that all known facts of electricity and magnetism (at that time all quasi-static) are consistent with the retarded potentials as much as with the static forms, he proceeds to derive equations for the fields that are the Maxwell equations we know, with an Ohm's law contribution for the assumed conducting medium. He points out that these equations are equivalent to those of his 1862 paper on light and proceeds to discuss light propagation and attenuation in metals, in dielectrics, in empty space, and the absence of free charge within conductors. He also works backward from the differential equations to obtain the retarded solutions for the potentials and the electric field in terms of the potentials in order to establish completely the equivalence of his theories of light and electromagnetism.

In the course of deriving his "Maxwell equations," Lorenz establishes that his retarded potentials are solutions of the wave equation and also must satisfy the condition, $d\bar{\Omega}/dt = -2(d\alpha/dx + d\beta/dy + d\gamma/dz)$ (*ibid.*, p. 294) or in modern notation and units,

$$\nabla \cdot \mathbf{A} + \frac{1}{c}\frac{\partial \Phi}{\partial t} = 0 \quad . \qquad (19)$$

This equation, now almost universally called the "Lorentz condition," is seen to originate with Lorenz more than 25 years before Lorentz. In discussing the quasi-static limit, Lorenz remarks (p. 292) that the retarded potentials (in modern, corrected terms, the "Lorenz" gauge potentials) give the same fields as the instantaneous scalar potential and a vector potential that is "a mean between Weber's and Neumann's theories," namely, (17), appropriate for Maxwell's choice of $\nabla \cdot \mathbf{A} = 0$. Without explicit reference, Lorenz was apparently aware of and made use of what we call gauge transformations.

Lorenz's paper makes no reference to Maxwell. Indeed, he only cites himself, but by 1868 Maxwell had read Lorenz's paper and in his *Treatise*, at the end of the chapter giving his electromagnetic theory of light, he mentions Lorenz's work as covering essentially the same ground (Maxwell, 1973, 1st ed., Note after Sect. 805, p. 398; 3rd ed., p. 449-450). Although Lorenz made a number of contributions to optics and electromagnetism during his career (Kragh, 1991, 1992), his pioneering papers were soon forgotten. A major contributing factor was surely Maxwell's objection to the retarded potentials of Riemann (1867) and Lorenz (1867):

> "From the assumptions of both these papers we may draw the conclusions, first, that action and reaction are not always equal and opposite, and second, that apparatus may be constructed to generate any amount of work from its resources." (Maxwell, 1868, *Sc. P.,* Vol. 2, p. 137).

Given the sanctity of Newton's third law and the conservation of energy, and Maxwell's stature, such criticism would be devastating. It is ironic that the person who almost invented electromagnetic momentum and who showed that all electromagnetic effects propagate with the speed of light did not recognize that the momentum of the electromagnetic fields needed to be taken into account in Newton's third law. Lorenz died in 1891, inadequately recognized then or later. In fact, by 1900 his name had disappeared from the mainstream literature on electromagnetism.

An interesting footnote on the Lorenz condition (19) is that in 1888, twenty years later, FitzGerald, trying to incorporate a finite speed of propagation into his mechanical "wheel and band" model of the ether, was bothered by the instantaneous character of Maxwell's scalar potential. His model would accommodate no such instantaneous behavior. Realizing that it was a consequence of $\nabla \cdot \mathbf{A} = 0$, he proposed (19) and found the standard wave equation of propagation for both $\Phi$ and $\mathbf{A}$. (Hunt, 1991, p. 115-118).

The mistaken attribution of (19) to Lorentz was pointed out by O'Rahilly (1938), Van Bladel (1991), and others. That Lorenz, not Lorentz, was the father of the retarded potentials (18) was first pointed out by Whittaker (1951, p.268) but he mistakenly states (*ibid.,* p.394) that Levi-Civita was the first to show (in 1897) that potentials defined by these integrals satisfy (19). Levi-Civita in fact does just what Lorenz did in 1867. Lorentz's own use of the Lorenz condition is discussed below.

Heinrich Hertz is most famous for his experiments in the 1880s demonstrating the free propagation of electromagnetic waves (Hertz, 1892), but he is equally important for his theoretical viewpoint. In 1884, beginning with the quasi-static, instantaneous electric and magnetic vector potentials of Helmholtz et al, he developed an iteration scheme that led to wave equations for the potentials and to the Maxwell equations in free space for the fields (Hertz, 1896, *Electric Waves,* p.273-290). His iterative approach showed one path from the action-at-a-distance potentials to the dynamical Maxwell equations for the fields. Hertz (*ibid.,* p.286) states that both

> "Riemann in 1858 and Lorenz in 1867, with a view to associating optical and electrical phenomena with one another, postulated the same or quite similar laws for the propagation of the potentials. These investigators recognized that these laws involve the addition of new terms to the forces which actually occur in electromagnetics; and they justify this by pointing out that these new terms are too small to be experimentally observable. But we see that the addition of these terms is far from needing any apology. Indeed their absence would necessarily involve contradiction of principles which are quite generally accepted."

It seems that Hertz did not fully appreciate that, while Lorenz's path from potentials to field equations was different in detail from his, Lorenz accomplished the same result 17 years earlier. Lorenz was not apologizing, but justifying his adoption of the retarded potentials as the necessary generalization, still in agreement with the known facts of electricity and magnetism. They were his starting point for obtaining his form of the Maxwell equations.

Six years later, Hertz (Hertz, 1892, p.193-268) addressed electrodynamics for bodies at rest and in motion. He discussed various applications, with the fields always to the fore and the scalar and vector potentials secondary. In this endeavor he made common cause with Heaviside, to whom he gives prior credit (Hunt, 1991, p. 122-128). Both men believed the potentials were unnecessary and confusing. In calculations Hertz apparently avoided them at all costs; Heaviside used them sparingly (O'Hara and Pricha, 1987, p.58, 62, 66-67). By using only the fields, Hertz avoided the issue of different forms of the potentials - his formalism was gauge invariant, by definition*.

___

* Hertz did not avoid potentials entirely. His name is associated with the "polarization potentials" of radiation problems.

___

### D. Charged particle dynamics - Clausius, Heaviside, and Lorentz (1892)

We have already described Weber's force equation (8) for the interaction of charged particles. While it permitted Weber to deduce the correct force between closed current-carrying circuits, it does not even remotely agree to order $1/c^2$ with the force between two charges in motion. It also implies inherently unphysical behavior, as shown by Helmholtz (1873). Weber's work was important nevertheless in its focus on charged particles instead of currents and its initiation of the Kirchhoff-Weber form of the vector potential.

A significant variation on charged particle dynamics, closer to the truth than Weber's, was proposed by Rudolf Julius Emanuel Clausius (1877, 1880). Struck by Helmholtz's demonstration of the equivalence of Weber's and Neumann's expressions for the interaction of charges or current elements, Clausius chose to write Lagrange's equations with an interaction of two charged particles $e$ and $e'$ that amounts to an interaction Lagrangian of the form,*

___

* Prior to the end of the 19th century in mechanics and the beginning of the 20th century in electrodynamics the compact notation of $L$ for $T - V$ was rarely used in writing Lagrange's equations. We use the modern notation $L_{int}$ as a convenient shorthand despite its absence in the papers cited.

___

$$L_{int} = \frac{ee'}{r}\left[-1 + \frac{\mathbf{v}\cdot\mathbf{v'}}{c^2}\right] \quad , \tag{20}$$

Generalized to one charge $e$ interacting with many, treated as continuous charge and current densities ($\rho$, $\mathbf{J}$), this Lagrangian reads

$$L_{int} = e\left[-\Phi(\mathbf{x}, t) + \frac{1}{c}\mathbf{v}\cdot\mathbf{A}_N(\mathbf{x}, t)\right] \quad, \tag{21}$$

where $\Phi$ is the instantaneous Coulomb potential and $\mathbf{A}_N$ is the instantaneous Neumann potential (7) with a time-dependent current. The interaction (20), inherent in Neumann's earlier work on currents, is a considerable step forward in the context of charged particle interactions, but its instantaneous action-at-a-distance structure means that it is not a true description, even to order $1/c^2$. (The force deduced from it has the correct magnetic field coupling to order $1/c^2$, but lacks some of the corresponding corrections to the electric field contribution.)

In an impressive paper, Oliver Heaviside (1889) chose $\nabla\cdot\mathbf{A} = 0$ (so that the instantaneous Coulomb field is exact) and constructed the appropriate vector potential, (17) for a point source, to give the velocity-dependent interaction correct to order $1/c^2$ (Heaviside, 1889, p.328, Eq.(8)). For two charges $e$ and $e'$ with velocities $\mathbf{v}$ and $\mathbf{v}'$, respectively, his results are equivalent to the interaction Lagrangian,

$$L_{int} = \frac{e\,e'}{r}\left[-1 + \frac{1}{2c^2}(\mathbf{v}\cdot\mathbf{v}' + \hat{\mathbf{r}}\cdot\mathbf{v}\,\hat{\mathbf{r}}\cdot\mathbf{v}')\right] \quad. \tag{22}$$

Heaviside also derived the magnetic part of the Lorentz force. His contributions, like Lorenz's, were largely ignored subsequently. Darwin (1920) derived (22) by another method with no reference to Heaviside and applied it to problems in the old quantum theory. See also Fock (1959).

A different approach was developed by H. A. Lorentz (1892) as part of his comprehensive statement of what we now call the microscopic Maxwell theory, with charges at rest and in motion as the sole sources of electromagnetic fields. His chapter IV is devoted to the forces between charged particles. The development is summarized on p.451-2 by statement of the microscopic Maxwell equations and the Lorentz force equation, $\mathbf{F} = e[\mathbf{E} + \mathbf{v}\times\mathbf{B}/c]$. In using D'Alembert's principle to derive his equations, Lorentz employs the vector potential, but never states explicitly its form in terms of the sources. It is clear, however, that he has retardation in mind, on the one hand from his exhibition of the full Maxwell equations to determine the fields caused by his $\rho$ and $\mathbf{J} = \rho\mathbf{v}$, and on the other by his words at the beginning of the chapter. He calls his reformulation (in translation from the French)
   "a fundamental law comparable to those of Weber and Clausius, while maintaining the consequences of Maxwell's principles."
A few sentences later, he stresses that the action of one charged particle on another is propagated at the speed of light, a concept originated by Gauss in 1845, but largely ignored for nearly 50 years.

Joseph Larmor (1900) used the principle of least action for the combined system of electromagnetic fields and charged particles to obtain both the Maxwell equations and the Lorentz force equation. Karl Schwarzschild, later renowned in astrophysics and general relativity, independently used the same technique to discuss the combined system of particles and fields (Schwarzschild, 1903). He was the first to write explicitly the familiar

Lagrangian $L_{int}$ describing the interaction of a charged particle $e$, with coordinate **x** and velocity **v**, with retarded external electromagnetic potentials,

$$L_{int} = e \left[ -\Phi(\mathbf{x}, t) + \frac{1}{c} \mathbf{v} \cdot \mathbf{A}(\mathbf{x}, t) \right] \quad , \tag{23}$$

where $\Phi$ and **A** are the potentials given by (18).

It is curious that, to the best of the authors' knowledge, the issue of gauge invariance of this charged particle Lagrangian did not receive general consideration in print until 1941 in the text by Landau and Lifshitz (1941). See also Bergmann (1946). The proof is simple. Under the gauge transformation (1a,b) the Lagrangian (23) is augmented by

$$\Delta L_{int} = e \left[ \frac{1}{c} \frac{\partial \chi}{\partial t} + \frac{1}{c} \mathbf{v} \cdot \nabla \chi \right] = \frac{e}{c} \frac{d\chi}{dt} \quad , \tag{24}$$

a total time derivative and so makes no contribution to the equations of motion. Perhaps this observation is too obvious to warrant publication in other than textbooks. We note that, in deriving the approximate Lagrangian attributed to him, Darwin (1920) expands the retarded potentials for a charged particle, which involve $\mathbf{r} = \mathbf{x}(t) - \mathbf{x}'(t')$, in powers of $(t'-t) = -r/c$, with coefficients of the primed particle's velocity, acceleration, etc., to obtain a tentative Lagrangian and then adds a total time derivative to obtain (22). Fock (1959) makes the same expansion, but then explicitly makes a gauge transformation to arrive at (22). These equivalent procedures exploit the arbitrariness of (24).

### E. Lorentz: the acknowledged authority, general gauge freedom

Our focus here is on how H. A. Lorentz became identified as the originator of both the condition (19) between $\Phi$ and **A** and the retarded solutions (18). In chapter VI, Lorentz (1892) presents, without attribution, a theorem that the integral

$$F(\mathbf{x}, t) = \frac{1}{4\pi} \int \frac{1}{r} s(x', t' = t - r/c) \, d^3x' \tag{25a}$$

is a solution of the inhomogeneous wave equation with $s(\mathbf{x}, t)$ as a source term,

$$\frac{1}{c^2} \frac{\partial^2 F}{\partial t^2} - \nabla^2 F = s(\mathbf{x}, t) \quad . \tag{25b}$$

He then uses such retarded solutions for time integrals of the vector potential in a discussion of dipole radiation.

In fact, the theorem goes back to Riemann in 1858 and Lorenz in 1861 and perhaps others. Riemann apparently read his paper containing the theorem to the Göttingen academy in 1858, but his death prevented publication, remedied only in Riemann (1867). In (Lorenz, 1867b), Lorenz states (25a,b) and remarks that the demonstration is easy, giving as reference his paper on elastic waves (Lorenz, 1861). It seems clear that in 1861 Lorenz was unaware of Riemann's oral presentation. The posthumous publication of Riemann's note occurred simultaneously with and adjacent to Lorenz's 1867 paper in Annalen*.

---

* Riemann (1867) showed that retardation led to the quasi-static instantaneous interactions of Weber and Kirchhoff, much as done by Lorenz (1867b), and remarked on the connection between the velocity of propagation of light and the ratio of electrostatic and electromagnetic units.

---

In (Lorentz, 1895, Sect. 32), Lorentz quotes the theorem (25a,b), citing (Lorentz, 1892) for proof, and then in Sect. 33 writes the components of a vector field in the form equivalent to (18) with $J = \rho v$. He does not call his vector field ($\psi_x$, $\psi_y$, $\psi_z$) the vector potential. Having obtained the wave equation for $H$ with $\nabla \times J$ as source term, he merely notes that if $H$ is defined as the curl of his vector field, it is sufficient that the field satisfy the wave equation with $J$ as source. We thus see Lorentz in 1895 explicitly exhibiting retarded solutions, but without the condition (19).

In a festschrift volume in honor of the 25th anniversary of Lorentz's doctorate, Emil Wiechert (1900) summarizes the history of the wave equation and its retarded solutions. He cites Riemann in 1858, Poincaré in 1891, Lorentz (1892, 1895), and Levi-Civita in 1897. No mention of Lorenz! In the same volume, des Coudres (1900) cites (Lorentz, 1892) for the theorem (25a,b) and calls the retarded solutions (18) "Lorentz'schen Lösungen." It is evident that by 1900 the physics community had attributed the retarded solutions for $\Phi$ and $A$ to Lorentz, to the exclusion of others.

Additional reasons for Lorentz being the reference point for modern classical electromagnetism are his magisterial encyclopedia articles (Lorentz, 1904a, 1904b), and his book (Lorentz, 1909). Here we find the first clear statement of the arbitrariness of the potentials under what we now call general gauge transformations. On p. 157 of (Lorentz, 1904b), he first states that in order to have the potentials satisfy the ordinary wave equations they must be related by

$$\nabla \cdot \mathbf{A} = -\frac{1}{c}\frac{\partial \Phi}{\partial t} \quad . \qquad \text{[Lorentz's (2)]}.$$

He then discusses the arbitrariness in the potentials, stating that other potentials $A_0$ and $\Phi_0$ may give the same fields, but not satisfy his constraint. He then states "every other admissible pair $A$ and $\Phi$" can be related to the first pair via the transformations,

$$\mathbf{A} = \mathbf{A}_0 - \nabla \chi \, , \quad \Phi = \Phi_0 + \frac{1}{c}\dot{\chi} \quad . \tag{26}$$

He then says that the scalar function $\chi$ can be found so that $A$ and $\Phi$ do satisfy [Lorentz's (2)] by solving the inhomogeneous wave equation,

$$\nabla^2 \chi - \frac{1}{c^2}\ddot{\chi} = \nabla \cdot \mathbf{A}_0 + \frac{1}{c}\dot{\Phi}_0 \quad . \tag{27}$$

A reader might question whether Lorentz was here stating the general principle of what we term gauge invariance. He stated his constraint before his statement of the arbitrariness of the potentials and then immediately restricted $\chi$ to a solution of (27). This doubt is removed in his book (Lorentz, 1909). There, in Note 5, he says,

"Understanding by $A_0$ and $\Phi_0$ special values, we may represent *other values that may as well be chosen by* [our equation (26)] *where χ is some scalar function* (emphasis added). We shall determine χ by subjecting $A$ and $\Phi$ to the condition [Lorentz's (2)] which can always be fulfilled because it leads to the equation [our equation (27)] which can be satisfied by a proper choice of χ."

He then proceeds to the wave equations and the retarded solutions in Sect. 13 of the main text. Lorentz obviously preferred potentials satisfying his constraint to the exclusion of other choices, but he did recognize the general principle of gauge invariance in classical electromagnetism without putting stress on it.

The dominance of Lorentz's publications as source documents is illustrated by their citation by G. A. Schott in his Adams Prize essay (Schott, 1912). On p. 4, Schott quotes (19) [his equation (IX)] and the wave equations for $A$ and $\Phi$. He then cites Lorentz's second Encyclopedia article (Lorentz, 1904b) and his book (Lorentz, 1909) for the retarded solutions (18) [his equations (X) and (XI)], which he later on the page calls "the Lorentz integrals."

Lorentz's domination aside, the last third of the 19th century saw the fundamentals of electromagnetism almost completely clarified, with the ether soon to disappear. Scientists went about applying the subject with confidence. They did not focus on niceties such as the arbitrariness of the potentials, content to follow Lorentz in use of the retarded potentials (18). It was only with the advent of modern quantum field theory and the construction of the electroweak theory and quantum chromodynamics that the deep significance of gauge invariance emerged.

## III. DAWNING OF THE QUANTUM ERA

### A. 1926 : Schrödinger, Klein, Fock

The year 1926 saw the flood gates open. Quantum mechanics, or more precisely, wave mechanics, blossomed at the hands of Erwin Schrödinger and many others. Among the myriad contributions, we focus only on those that relate to our story of the emergence of the principle of gauge invariance in quantum theory. The pace among this restricted set is frantic enough. [To document the pace, we augment the references for the papers in this era with submission and publication dates.] The thread we pursue is the relativistic wave equation for spinless charged particles, popularly known nowadays as the Klein-Gordon equation. The presence of both the scalar and vector potentials brought forth the discovery of the combined transformations (1 a,b,c) by Fock.

The relativistic wave equation for a spinless particle with charge $e$ interacting with electromagnetic fields is derived in current textbooks by first transforming the classical constraint equation for a particle of 4-momentum $p^\mu = (p^0, \boldsymbol{p})$ and mass $m$, $p^\mu p_\mu = (mc)^2$, by the substitution $p^\mu \to p^\mu - eA^\mu/c$, where $A^\mu = (A^0 = \Phi, \boldsymbol{A})$ is the 4-vector electromagnetic potential. Here we use the metric $g^{00} = 1$, $g^{ij} = -\delta_{ij}$. Then a quantum mechanical operator acting on a wave function $\psi$ is constructed by the operator substitution, $p^\mu \to i\hbar\partial^\mu$, where $\partial^\mu = \partial/\partial x_\mu = (\partial^0, -\boldsymbol{\nabla})$. Explicitly, we have

$$(i\hbar\partial^\mu - eA^\mu/c)(i\hbar\partial_\mu - eA_\mu/c)\,\psi = (mc)^2\,\psi \quad . \tag{28}$$

Alternatively, we divide through by $-\hbar^2$ and write

$$\left[ (\partial^\mu + ie A^\mu/\hbar c)(\partial_\mu + ie A_\mu/\hbar c) + (mc/\hbar)^2 \right] \psi = 0 \quad . \tag{29}$$

Separation of the space and time dimensions and choice of a constant energy solution, $\psi \propto \exp(-iEt/\hbar)$, yields the relativistic version of the Schrödinger equation,

$$-\hbar^2 c^2 \nabla^2 \psi + ie\hbar c (\partial^\mu A_\mu) \psi + 2ie\hbar c \mathbf{A}\cdot\nabla\psi + e^2 \mathbf{A}\cdot\mathbf{A}\,\psi = \left[ (E - e\Phi)^2 - (mc^2)^2 \right] \psi \quad . \tag{30}$$

The second term on the left is absent if the Lorenz gauge condition $\partial^\mu A_\mu = 0$ is chosen for the potentials.

    The first of Schrödinger's four papers (Schrödinger, 1926a) was submitted on 27 January and published on 13 March. It was devoted largely to the nonrelativistic time-independent wave equation and simple potential problems, but in Section 3 he mentions the results of his study of the "relativistic Kepler problem." (An English translation of Schrödinger's 1926 papers can be found in (Schrödinger, 1978)). According to his biographer (Moore, 1989, p.194-197), Schrödinger derived the relativistic wave equation in November 1925, began solving the problem of the hydrogen atom while on vacation at Christmas, and completed it in early January 1926. Disappointed that he had not obtained the Sommerfeld fine-structure formula, he did not publish his work, but focused initially on the nonrelativistic equation. Some months later, in Sect. 6 of his fourth paper (Schrödinger, 1926b), he tentatively presented the relativistic equation in detail and discussed its application to the hydrogen atom and to the Zeeman effect.

    Schrödinger was not the only person to consider the relativistic wave equation. In a private letter to Jordan dated 12 April 1926, Pauli used the relativistic connection between energy and momentum to derive a wave equation equivalent to (29) with a static potential, then specialized to the nonrelativistic Schrödinger equation, and went on to his main purpose - to show the equivalence of matrix mechanics and wave mechanics (van der Waerden, 1973). In the published literature, Oskar Klein (1926) treated a five-dimensional relativistic formalism and explicitly exhibited the four-dimensional relativistic wave equation for fixed energy with a static scalar potential. He showed that the nonrelativistic limit was the time-independent Schrödinger equation, but did not discuss any solutions. Before publication of Klein's paper, Fock (1926a) independently derived the relativistic wave equation from a variational principle and solved the relativistic Kepler problem. He observed that Schrödinger had already commented on the solution in his first paper. In his paper Fock did not include the general electromagnetic interaction. Schrödinger comments in the introduction ("Abstract") to his collected papers (Schrödinger, 1978),

> "V. Fock carried out the calculations quite independently in Leningrad, before my last paper [(Schrödinger, 1926b)] was sent in, and also succeeded in deriving the relativistic equation from a variational principle. *Zeitschrift für Physik* **38**, 242 (1926)."

    The discovery of the symmetry under gauge transformations (1 a,b,c) of the quantum mechanical system of a charged particle interacting with electromagnetic fields is due to Fock (1926b). His paper was submitted on 30 July 1926 and published on 2 October 1926. In it he first discussed the special-relativistic wave equation of his earlier paper with electromagnetic interactions and addressed the effect of the change in the potentials (1 a,b). He showed that

the equation is invariant under the change in the potentials provided the wave function is transformed according to (1c). He went on to treat a five-dimensional general-relativistic formalism, similar to but independent of Klein. In a note added in proof, Fock notes that

> "While this note was in proof, the beautiful work of Oskar Klein [published on 10 July] arrived in Leningrad,"

and that the principal results were identical.

That fall others contributed. Kudar (1926) wrote the relativistic equations down in covariant notation, citing Klein (1926) and Fock (1926a). He remarked that his general equation reduced to Fock's for the Kepler problem with the appropriate choice of potentials. Walter Gordon (1926) discussed the Compton effect using the relativistic wave equation to describe the scattering of light by a charged particle. He referred to Schrödinger's first three papers, but not the fourth (Schrödinger, 1926b) in which Schrödinger actually treats the relativistic equation. Gordon does not cite Klein or either of Fock's papers.

The above paragraphs show the rapid pace of 1926, the occasional duplication, and the care taken by some, but not all, for proper acknowledgment of prior work by others. If we go chronologically by publication dates, the Klein-Gordon equation should be known as the Klein-Fock-Schrödinger equation; if by notebooks and letters, Schrödinger and Pauli could claim priority. Totally apart from the name attached to the relativistic wave equation, the important point in our story is Fock's paper on the gauge invariance, published on 2 October 1926 (Fock, 1926b).

The tale now proceeds to the enshrinement by Weyl of symmetry under gauge transformations as a guiding principle for the construction of a quantum theory of matter (electromagnetism and gravity). Along the way, we retell the well-known story of how the seemingly inappropriate word "gauge" came to be associated with the transformations (1 a,b,c) and today's generalizations.

## B. Weyl: gauge invariance as a basic principle

Fritz London, in a short note in early 1927 (London, 1927a) and soon after in a longer paper (London, 1927b), proposed a quantum mechanical interpretation of Weyl's failed attempt to unify electromagnetism and gravitation (Weyl, 1919). This attempt was undertaken long before the discovery of quantum mechanics. London noticed that Weyl's principle of invariance of his theory under a scale change of the metric tensor $g^{\mu\nu} \to g^{\mu\nu} exp\, \lambda(x)$, where $\lambda(x)$ is an arbitrary function of the space-time coordinates, was equivalent in quantum mechanics to the invariance of the wave equation under the transformations (1) provided $\lambda(x)$ was made imaginary. In his short note London cites Fock (1926b) but does not repeat the citation in his longer paper, although he does mention (without references) both Klein and Fock for the relativistic wave equation in five dimensions.

To understand London's point we note first that Weyl's incremental change of length scale $d\ell = \ell\, \phi_\nu dx^\nu$ leads to a formal solution $\ell = \ell_0 exp\lambda(x)$, where $\lambda(x) = \int^x \phi_\nu dx^\nu$; the indefinite integral over the real "potential" $\phi_\nu$ is path-dependent. If we return to the relativistic wave equation (29), we observe that a *formal* solution for a particle interacting with the electromagnetic potential $A^\mu$ can be written in terms of the solution without interaction as

$$\psi = exp\left(-i\frac{e}{\hbar c}\int^{x} A_{\nu}dx^{\nu}\right)\psi_0 \quad , \tag{31}$$

where $\psi_0$ is the zero-field solution. [Recovery of (29) may be accomplished by "solving" for $\psi_0$ and requiring $\partial^{\mu}\partial_{\mu}\psi_0 + (mc/\hbar)^2\psi_0 = 0$.] With the gauge transformation of the 4-vector potential

$$A^{\mu} \rightarrow A'^{\mu} = A^{\mu} - \partial^{\mu}\chi \quad , \tag{32}$$

the difference in phase factors is obviously the integral of a perfect differential, $-\partial_{\mu}\chi dx^{\mu}$. Up to a constant phase, the wave functions $\psi'$ and $\psi$ are thus related by the phase transformation

$$\psi' = exp(ie\chi(x)/\hbar c)\,\psi \quad , \tag{33}$$

which is precisely Fock's (1c). London actually expressed his argument in terms of "scale change" $\ell = \ell_0 \exp(i\lambda(x))$, where $i\lambda(x)$ is the quantity in the exponential in (31), and wrote $\psi/\ell = \psi_0/\ell_0$.

The "gradient invariance" of Fock became identified by London and then by Weyl with an analogue of Weyl's "eichinvarianz" (scale invariance), even though the former concerns a local phase change and the latter a coordinate scale change. In his famous book, "Gruppentheorie und Quantenmechanik" (Weyl, 1928), Weyl discusses the coupling of a relativistic charged particle with the electromagnetic field. He observes, without references, that the electromagnetic equations and the relativistic Schrödinger equation (28) are invariant under the transformations (1a,b,c). Weyl then states on p. 88 (in translation):

> "This *'principle of gauge invariance'* is quite analogous to that previously set up by the author, on speculative grounds, in order to arrive at a unified theory of gravitation and electricity[22]. But I now believe that this gauge invariance does not tie together electricity and gravitation, but rather *electricity and matter* in the manner described above."

His note 22 refers to his own work, to Schrödinger (1923), and to London (1927b). In the first (1928) edition, the next sentence reads (again in translation):

> "How gravitation according to general relativity must be incorporated is not certain at present."

By the second (1931) edition, this sentence has disappeared, undoubtedly because he believed that his own work in the meantime (Weyl, 1929a, 1929b) had shown the connection. In fact, in the second edition a new section 6 appears in Chapter IV, in which Weyl elaborates on how the gauge transformation (1c) can only be fully understood in the context of general relativity.

Weyl's 1928 book and his papers in 1929 demonstrate an evolving point of view unique to him. Presumably prompted by London's observation, he addressed the issue of gauge invariance in relativistic quantum mechanics, knowing on the one hand that the principle obviously applied to the electromagnetic fields and charged matter waves, and on the other hand wanting to establish contact with his 1919 "eichinvarianz." As we have just seen, in his 1928 book he presented the idea of gauge invariance in the unadorned version of Fock, without the "benefit" of general relativity. But in the introduction of the first of his 1929 papers on the electron and gravitation (Weyl, 1929a), he states the "principle of gauge invariance" (the first use of the words in English) very much as in his book, citing only it for authority. He then goes on to show that the conservation of electricity is a double consequence of gauge invariance (through the matter and the electromagnetic equations) and that

> "This new principle of gauge invariance, which may go by the same name, has the character of general relativity since it contains an arbitrary function λ, and can certainly only be understood with reference to it."

He elaborated on this point in (Weyl, 1929b):

> " In special relativity one must regard this gauge-factor as a constant because here we have only a single point-independent tetrad. Not so in general relativity; every point has its own tetrad and hence its own arbitrary gauge-factor: because by the removal of the rigid connection between tetrads at different points the gauge-factor necessarily becomes an arbitrary function of position." (translation taken from O'Raifeartaigh and Straumann, 2000, p. 7).

Nevertheless, Weyl stated (Weyl, 1929a, p.332, below equation (8)),

> "If our view is correct, then the electromagnetic field is a necessary accompaniment of the matter wave field and not of gravitation."

The last sentence of (Weyl, 1929b) contains almost the same words. His viewpoint about the need for general relativity can perhaps be understood in the sense that λ *must be* an arbitrary function in the curved space-time of general relativity, but not necessarily in special relativity, and his desire to provide continuity with his earlier work. The close mathematical relation between non-abelian gauge fields and general relativity as connections in fiber bundles was not generally realized until much later (See e.g., Yang, 1986; O'Raifeartaigh and Straumann, 2000).

Historically, of course, Weyl's 1929 papers were a watershed. They enshrined as fundamental the modern principle of gauge invariance, in which the existence of the 4-vector potentials (and field strengths) follow from the requirement of the invariance of the matter equations under gauge transformations such as (1c) of the matter fields. This principle is the touchstone of the theory of gauge fields, so dominant in theoretical physics in the second half of the 20th century. The important developments beyond 1929 can be found in the reviews already mentioned in the Introduction. The reader should be warned, however, of a curiosity regarding the citation of Fock's 1926 paper (Fock, 1926b) by O'Raifeartaigh (1997), O'Raifeartaigh and Straumann (2000), and Yang (1986, 1987). While the volume and page number are given correctly, the year is invariably given as 1927. One of the writers privately blames it on Pauli (1933). Indeed, Pauli made that error, but he did give to Fock the priority of introducing gauge invariance in quantum theory.

## IV. PHYSICAL MEANING OF GAUGE INVARIANCE, EXAMPLES

### A. On the physical meaning of gauge invariance in QED and quantum mechanics

While for Electroweak Theory and QCD gauge invariance is of paramount importance, its physical meaning in QED per se does not seem to be extremely profound. A tiny mass of the photon would destroy the gauge invariance of QED, as a mass term $m_\gamma^2 A^2$ in the Lagrangian is not gauge invariant. At the same time the excellent agreement of QED with experiment and in particular its renormalizability would not be impaired (see e.g., Kobzarev and Okun, 1968; Goldhaber and Nieto, 1971). On the other hand the renormalizability would be destroyed by an anomalous magnetic moment term in the Lagrangian, $\mu\bar{\psi}\sigma_{\lambda\nu}\psi F^{\lambda\nu}$, in spite of its manifest gauge invariance. What is really fundamental in electrodynamics is the conservation of electromagnetic current or in other words conservation of charge (see e.g., Okun, 1986, Lecture 1). Conservation of charge makes the effects caused by a possible nonvanishing mass of the photon, $m_\gamma$, proportional to $m_\gamma^2$ and therefore negligibly small for small enough values of $m_\gamma$.

It should be stressed that the existing upper limits on the value of $m_\gamma$ lead in the case of non-conserved current to such catastrophic bremsstrahlung, that most of the experiments which search for monochromatic photons in charge-nonconserving processes become irrelevant (Okun and Zeldovich, 1978). Further study has shown (Voloshin and Okun, 1978) that reabsorption of virtual bremsstrahlung photons restores the conservation of charge (for reviews, see Okun, 1989, 1992).

As has been emphasized above, gauge invariance is a manifestation of non-observability of $A^\mu$. However integrals such as in eq (31) are observable when they are taken over a closed path, as in the Aharonov-Bohm effect (Aharonov and Bohm, 1959). The loop integral of the vector potential there can be converted by Stokes's theorem into the magnetic flux through the loop, showing that the result is expressible in terms of the magnetic field, albeit in a nonlocal manner. It is a matter of choice whether one wishes to stress the field or the potential, but the local vector potential is not an observable.

## B. Examples of gauges

The gauge invariance of classical field theory and of electrodynamics in particular allows one to consider the potential $A^\mu$ with various gauge conditions, most of them being not Poincaré invariant:

$$\partial_\mu A^\mu = 0 \ (\mu = 0, 1, 2, 3), \quad \text{Lorenz gauge} \tag{34}$$

$$\mathbf{\nabla}\cdot\mathbf{A} = \partial_j A_j = 0 \ (j = 1, 2, 3), \quad \text{Coulomb gauge or radiation gauge} \tag{35}$$

$$n_\mu A^\mu = 0 \ (n^2 = 0), \quad \text{light cone gauge} \tag{36}$$

$$A_o = 0, \quad \text{Hamiltonian or temporal gauge} \tag{37}$$

$$A_3 = 0, \quad \text{axial gauge} \tag{38}$$

$$x_\mu A^\mu = 0, \quad \text{Fock-Schwinger gauge} \tag{39}$$

$$x_j A_j = 0, \quad \text{Poincaré gauge} \tag{40}$$

An appropriate choice of gauge simplifies calculations. This is illustrated by many examples presented in textbooks, e.g., (Jackson, 1998). For the quantum mechanics of nonrelativistic charged particles interacting with radiation, the Coulomb gauge is particularly convenient because the instantaneous scalar potential describing the static interactions and binding is unquantized; only the transverse vector potential of the photons is quantized. However, noncovariant gauges, characterized by fixing a direction in Minkowski space, pose a number of problems discussed in Gaig, Kummer, and Schweda (1990). The problems acquire additional dimensions in quantum field theory where one has to deal with a space of states and with a set of operators.

In QED the gauge degree of freedom has to be fixed before the theory is quantized. Usually the gauge fixing term $(\partial_\mu A^\mu)^2$ is added to the gauge invariant Lagrangian density with coefficient $1/2\alpha$ (For futher details see Gaig, Kummer, and Schweda, 1990; Berestetskii, Lifshitz, and Pitaevskii, 1971; Ramond, 1981; Zinn-Justin, 1993). In perturbation theory the propagator of a virtual photon with 4-momentum $k$ acquires the form,

$$D(k)^{\mu\nu} = -\frac{1}{k^2}\left[g^{\mu\nu} + (\alpha - 1)\frac{k^\mu k^\nu}{k^2}\right] \quad . \tag{41}$$

The most frequently used cases are

$$\alpha = 1 \quad (\text{Feynman gauge}), \tag{42}$$

$$\alpha = 0 \quad (\text{Landau gauge}). \tag{43}$$

In the Feynman gauge the propagator (41) is simpler, while in the Landau gauge its longitudinal part vanishes, which is often more convenient. If calculations are carried out correctly, the final result will not contain the gauge parameter $\alpha$.

In the static (zero frequency) limit the propagator (41) reduces to

$$D_{ij}(\mathbf{k}, 0) = \frac{1}{|\mathbf{k}|^2}\left(\delta_{ij} + (\alpha - 1)\frac{k_i k_j}{|\mathbf{k}|^2}\right) \quad , \tag{44}$$

$$D_{00}(\mathbf{k}, 0) = \frac{1}{|\mathbf{k}|^2} \quad , \tag{45}$$

$$D_{0j}(\mathbf{k}, 0) = D_{i0}(\mathbf{k}, 0) = 0 \quad . \tag{46}$$

This is the propagator for the Helmholtz potential (14) and the static Coulomb potential.

Various gauges have been associated with names of physicists, a process begun by Heitler who introduced the term "Lorentz relation" in the first edition of his book (Heitler, 1936). In the third edition (Heitler, 1954) he used "Lorentz gauge" and "Coulomb gauge." Zumino (1960) introduced the terms "Feynman gauge," "Landau gauge," and "Yennie gauge" ($\alpha = 3$ in (41)).

## V.  SUMMARY AND CONCLUDING REMARKS

What is now generally known as a gauge transformation of the electromagnetic potentials (1a, b) was discovered in the process of formulation of classical electrodynamics by its creators, Lorenz, Maxwell, Helmholtz, and Lorentz, among others (1867-1909). The phase transformation (1c) of the quantum mechanical charged field accompanying the transformation of the electromagnetic potentials was discovered by Fock (1926b). The term "gauge" was applied to this transformation by Weyl (1928, 1929a, 1929b) (who used "eich-" a decade before to denote a *scale* transformation in his unsuccessful attempt to unify gravity and electromagnetism).

In text books on classical electrodynamics the gauge invariance (1a,b) was first discussed by Lorentz in his influential book, "Theory of Electrons" (Lorentz, 1909). The first derivation of the invariance of the Lagrangian for the combined system of electromagnetic fields and charged particles was presented by Landau and Lifshitz (1941) (with reference to Fock, they used Fock's term "gradient invariance").

The first model of a non-abelian gauge theory of weak, strong, and electromagnetic interactions was proposed by Klein (1938) (who did not use the term "gauge" and did not refer to Weyl). But this attempt was firmly forgotten. The modern era of gauge theories started with the paper by Yang and Mills (1954).

The history of gauge invariance resembles a random walk, with the roles of some important early players strangely diminished with time. There is a kind of echo between the loss of interest by O. Klein and L. Lorenz in their "god blessed children". It is striking that the notion of gauge symmetry did not appear in the context of classical electrodynamics, but required the invention of quantum mechanics. It is amusing how little the authors of text-books know about the history of physics. For a further reading the Resource Letter (Cheng and Li, 1988) is recommended.

## ACKNOWLEDGEMENTS

The authors thank Robert N. Cahn, David J. Griffiths, Helge Kragh, and Valentine L. Telegdi for their assistance and advice. The work of JDJ was supported in part by the Director, Office of Science, Office of High Energy and Nuclear Physics, of the U.S. Department of Energy under Contract DE-AC03-76SF00098. The work of LBO was supported in part by grant RFBR # 00-15-96562, by an Alexander von Humboldt award, and by the Theory Division, CERN.

27# REFERENCES

Aharonov, Y., and D. Bohm, 1959, "Significance of Electromagnetic Potentials in the Quantum Theory," Phys. Rev. **115**, 485-491.

Ampère, A.-M., 1827, "Sur la théorie mathématique des phénomèns électrodynamiques uniquement déduite de l'expérience," Mémoires de l'Académie Royale des Sciences de l'Institut de France, ser. 2, **6**, 175-388 [memoirs presented from 1820 to 1825].

Berestetskii, V. B., E. M. Lifshitz, and L. P. Pitaevskii, 1971, *Relativistic Quantum Theory*, *Part 1* ( Pergamon, Oxford), Section 77.

Bergmann, P. G., 1946, *Introduction to the Theory of Relativity* (Prentice-Hall, New York), p.115-117.

Biot, J. B., and F. Savart, 1820, "Expériences électro-magnétiques," Journal de Phys., de Chimie, **91**, 151; "Note sur le Magnétisme de la pile de Volta," Ann. de Chimie et de Phys, ser. 2, **15,** 222-223.

Biot, J. B., 1824, *Précis Elémentaire de Physique Expérimentale,* 3rd ed., vol 2 (Deterville, Paris), p. 745.

Bork, A. M, 1967, "Maxwell and the Vector Potential," Isis **58**, 210-222.

Buchwald, J. Z., 1985, *From Maxwell to Microphysics* (University of Chicago Press, Chicago)

Buchwald, J. Z., 1989, *The Rise of the Wave Theory of Light* (University of Chicago Press, Chicago)

Buchwald, J. D., 1994,*The Creation of Scientific Effects, Heinrich Hertz and Electric Waves* (University of Chicago Press, Chicago)

Cheng, T. P., and Ling-Fong Li, 1988, "Resource Letter: GI-1 Gauge invariance," Am. J. Phys. **56**, 586-600.

Clausius, R., 1877, "Ueber die Ableitung eines neuen elektrodynamischen Grundgestezes," Journal für Math. (Crelle's Journal) **82**, 85-130.

Clausius, R., 1880, "On the employment of the electrodynamic potential for the determination of the ponderomotive and electromotive forces," Phil. Mag., ser. 5, **10**, 255-279 des Coudres, Th., 1900, "Zur Theorie des Kraftfeldes elektrischer Ladungen, die sich mit Ueberlichtgeschwindigkeit Bewegen," Arch. Néerl. Scs., ser. 2, **5**, 652-664.

Darrigol, O., 2000, *Electrodynamics from Ampère to Einstein* (Oxford University Press)

Darwin, C. G., 1920, "The Dynamical Motions of Charged particles," Phil. Mag., ser. 6, **39**, 537-551.

Everitt, C. W. F., 1975, *James Clerk Maxwell, Physicist and Natural Philosopher* (Scribner's, New York)

Faraday, M., 1839, *Experimental Researches in Electricity*, Vol. 1 (Richard & John Edward Taylor, London), p.1-41. [from Phil. Trans. Roy. Soc., November 24, 1831]

Fock, V., 1926a, "Zur Schrödingerschen Wellenmechanik," Zeit. für Phys. **38**, 242-250. [subm. 11 June 1926, publ. 28 July 1926]

Fock, V., 1926b, "Über die invariante Form der Wellen- und der Bewegungsgleichungen für einen geladenen Massenpunkt," Zeit. für Physik **39**, 226-232.[subm. 30 July 1926, publ. 2 October 1926]

Fock, V., 1959, *Theory of Space, Time and Gravitation*, transl. M. Hamermesh, (Pergamon, London), 2nd ed. (1965), Sect. 26.

Gaig, P., W. Kummer, and M. Schweda, 1990, Eds., *Physical and Nonstandard Gauges*, Proceedings of a Workshop, Vienna, Austria (September 19-23, 1989), Lecture Notes in Physics, 361 (Springer, Berlin-Heidelberg).

Grassmann, H., 1845, "Neue Theorie der Electrodynamik," Ann. der Phys. und Chem., **64**, 1-18.

Goldhaber, A. S., and M. M. Nieto, 1971, "Terrestrial and extraterrestrial limits on photon mass," Rev. Mod. Phys. **43**, 277-296.

Gordon, W., 1926, "Der Comptoneffekt nach der Schrödingerschen Theorie," Zeit. für Phys. **40**, 117-133. [subm, 29 September 1926, publ. 29 November 1926]

Heaviside, O., 1889, "On the Electromagnetic Effects due to the Motion of Electrification through a Dielectric," Phil. Mag. ser. 5, **27**, 324-339.

Heitler, W., 1936, *Quantum Theory of Radiation*, 1st ed.(Oxford University Press)

Heitler, W., 1954, *Quantum Theory of Radiation*, 3rd ed.(Oxford University Press)

Helmholtz, H., 1870, "Ueber die Bewegungsgleichungen der Elektricität für ruhende leitende Körper," Journal für die reine und angewandte Mathematik **72**, 57-129.

Helmholtz, H., 1872, "On the Theory of Electrodynamics," Phil. Mag. ser. 4, **44**, 530-537.

Helmholtz, H., 1873, "Ueber die Theorie der Elektrodynamik," Journal für die reine und angewandte Mathematik **75**, 35-66.[called Zweite Abhandlung Kritisches]

Helmholtz, H., 1874, "Ueber die Theorie der Elektrodynamik," Journal für die reine und angewandte Mathematik **78**, 273-324.[called the Dritte Abhandlung, and subtitled "Die elektrodynamischen Kräfte in bewegten Leitern."]

Hertz, H., 1892, *Untersuchungen ueber die Ausbreitung der elektrischen Kraft* (J. A. Barth, Leipzig); transl., *Electric Waves,* Auth. English translation by D. E. Jones (Macmillan, London, 1893; reprinted, Dover Publications, New York, 1962)

Hertz, H., 1896, *Miscellaneous Papers*, Auth. English translation by D. E. Jones and G. A. Schott (Macmillan, London).

Hunt, B. J., 1991, *The Maxwellians* (Cornell University Press)

Jackson, J. D., 1998, *Classical Electrodynamics,* 3rd ed. (Wiley, New York)

Jelved, K., A. D. Jackson, and O. Knudsen, 1998, Transl. & Eds., *Selected Scientific Works of Hans Christian Oersted* (Princeton University Press), p. 413-420.

Kirchhoff, G., 1857, "II. Ueber die Bewegung der Elektricität in Leitern," Annalen der Physik und Chemie **102**, 529-544. [reprinted in *Gesammelte Abhandlungen von G. Kirchhoff* (J. A. Barth, Leipzig 1882), p. 154-168]

Klein, O., 1926, "Quantentheorie und fünfdimensionale Relativitätstheorie," Zeit. für Phys. **37**, 895-906. [subm. 28 April 1926, publ. 10 July 1926]

Klein, O., 1938, "On the Theory of Charged Fields," in *New Theories in Physics*, Conference organized in collaboration with the International Union of Physics and the Polish Intellectual Cooperation Committee (Warsaw, May 30th- June 3rd, 1938).

Kobzarev, I. Yu., and L. B. Okun, 1968, "On the photon mass," Usp. Fiz. Nauk **95**, 131-137 [Sov. Phys. Uspekhi **11**, 338-341.]

Kragh, H., 1991, "Ludvig Lorenz and nineteenth century optical theory: the work of a great Danish scientist," Appl. Optics **30**, 4688-4695.

Kragh, H., 1992, "Ludvig Lorenz and the Early Theory of Long-distance Telephony," Centaurus **35**, 305-324.

Kudar, J., 1926, "Zur vierdimensionale Formulierung der undulatorischen Mechanik," Ann. der Physik **81**, 632-636. [subm. 30 August 1926, publ. 26 October 1926]

Landau, L. D., and E. M. Lifshitz, 1941, *Teoria Polya*, GITTL M.-L., Sect. 16; transl., *Classical Theory of Fields*, rev. 2nd ed. (Pergamon, Oxford 1962), Sect. 18.

Larmor J., 1900, *Ether and Matter* (Cambridge University Press), Ch. VI, esp. Sects. 56-58.

London, F., 1927a, "Die Theorie von Weyl und die Quantenmechanik," Naturwiss. **15**, 187. [subm. 19 January 1927, publ. 25 February 1927]

London, F., 1927b, "Quantenmechanische Deutung der Theorie von Weyl," Zeit. für Physik **42**, 375-389. [subm. 25 February 1927; publ. 14 April 1927]

Lorentz, H. A., 1892, "La théorie électromagnétique de Maxwell et son application aux corps mouvants," Arch. Néerl. Scs. **25**, 363-552.

Lorentz, H. A., 1895, *Versuch einer Theorie der Electrischen und Optischen Erscheinungen in begwegten Körpern* ( E. J. Brill, Leiden).

Lorentz, H. A., 1904a, Encykl. Math. Wissen., Band V:2, Heft 1, V. 13, "Maxwell's Elektromagnetische Theorie," p. 63-144.

Lorentz, H. A., 1904b, *ibid.*, V. 14, "Weiterbildung der Maxwellischen Theorie. Elektronentheorie," p. 145-280.

Lorentz, H. A., 1909, *Theory of Electrons* (G. B. Teubner, Leipzig and G. E. Stechert, New York).[2nd ed., 1916; reprinted by Dover Publications, New York. 1952]

Lorenz, L. V., 1861, "Mémoire sur la théorie de l'elasticité des corps homogènes à élasticité constante," J. für Math. (Crelle's Journal) **58**, 329-351.

Lorenz, L. V., 1863, "Die Theorie des Lichtes, I," Annalen der Physik und Chemie **18**, 111-145; "On the Theory of Light," Phil. Mag. ser. 4, **26**, 81-93, 205-219.

Lorenz, L. V., 1867a, "Om lyset," Tidsckr. Fys. Chem. **6**, 1-9.

Lorenz, L. V., 1867b, "Ueber die Identität der Schwingungen des Lichts mit den elektrischen Strömen," Ann. der Physik und Chemie **131**, 243-263; "On the Identity of the Vibrations of Light with Electrical Currents, "Phil. Mag. ser. 4, **34**, 287-301.

MacCullagh, J., 1839, "Dynamical Theory of crystalline Reflexion and Refraction," Proc. Roy. Irish Acad. **1**, No,. 20, 374-379; Trans. Roy. Irish Acad. **21**, part 1, 17-50 (1846); *Collected Works of James MacCullagh*, eds. J. H. Jellet and S. Haughton (Hodges, Figgis, Dublin, 1880). p.145 - 184.

Maxwell, J. C., 1856, "On Faraday's Lines of Force," Camb. Phil. Trans. **10**, 27-83; *Scientific Papers* (Dover reprint), Vol. 1, p. 155-229. [p. 188-229 is Part II, "On Faraday's 'Electro-tonic State'."]

Maxwell, J. C., 1865, "A Dynamical Theory of the Electromagnetic Field," Phil. Trans. Roy. Soc. **155,** 459-512; *Scientific Papers* (Dover reprint), Vol. 1, p. 526-597. [Part VI is on the electromagnetic theory of light.]

Maxwell, J. C., 1868, "On a Method of Making a Direct Comparison of Electrostatic and Electromagnetic Force; with a Note on the Electromagnetic Theory of Light," Phil. Trans. Roy. Soc. **158**, 643-657; *Scientific Papers* (Dover reprint), Vol. 2, p. 125-143.

Maxwell, J. C., 1873, *Treatise on Electricity and Magnetism*, 1st ed. (Clarendon Press, Oxford), Vol. II [3rd edition, 1891, Dover Publications, New York, 1954]

Moore, W. J., 1989, *Schrödinger, life and thought* (Cambridge University Press)

Neumann, F. E., 1847, "Allgemeine Gesetze der inducirten elektriche Ströme," Abhandlungen der Königlichen Akademie der Wissenschaften zu Berlin, aus dem Jahre 1845, 1-87.

Neumann, F. E., 1849, "Über ein allgemeines Princip der mathematischen Theorie inducirter elektricher Ströme," Abhandlungen der Königlichen Akademie der Wissenschaften zu Berlin, aus dem Jahre 1847, 1-72.

Novozhilov, Y. V., and V. Y. Novozhilov, 1999, "Works of Vladimir Aleksandrovich Fock in quantum field theory (On the centennial of the birth of V. A. Fock)," Teor. Mat. Fiz. **120**, 400-416 [Theor. Math. Phys **120**, 1150-1163].

Novozhilov, Y. V., and V. Y. Novozhilov, 2000, "Vladimir A. Fock (dedicated to the 100th anniversary of birth of academician V.A. Fock)," Fiz. Elem. Cha. Atom. Yad. **31**, 5-46. [Phys. Part. Nucl. **31**, 1-21].

O'Hara, J. G. and W. Pricha, 1987, *Hertz and the Maxwellians* (Peter Peregrinus, London)

Okun, L. B., and Ya. B. Zeldovich, 1978, "Paradoxes of unstable electron," Phys. Lett. **78B**, 597-599.

Okun, L. B., 1986, "Special topics on gauge theories," Surveys in High Energy Physics **5**, no.3, Five lectures 199-235, Supplement 236-285. [Originally published in 1983 JINR-CERN School of Physics, Tabor, Czechoslovakia, 5-18 June 1983, vol. 2 (Dubna 1984), pp.2-81.]

Okun, L. B., 1989, "Tests of electric charge conservation and the Pauli principle," Usp. Fiz. Nauk **158**, 293-301. [Sov. Phys. Uspekhi **32**, 543-547].

Okun, L. B., 1992, "Note on testing charge conservation and the Pauli exclusion principle," Phys. Rev. D **45**, No.11, part II, VI-10 - VI.14.

O'Rahilly, A., 1938, *Electromagnetics* (Longmans, Green and Cork University Press), reprinted as *Electromagnetic Theory* (Dover Publications, N.Y. 1965), footnote, p.184.

O'Raifeartaigh, L., 1997, *The Dawning of Gauge Theory* (Princeton University Press, Princeton, NJ). [A reprint volume with commentary].

O'Raifeartaigh, L., and N. Straumann, 2000, "Gauge theory: Historical origins and some modern developments," Rev. Mod. Phys. **72**, 1-23.

Pauli, W., 1933, in H. Geiger and K. Scheel, *Handbuch der Physik, Vol. XXIV/1, Quantentheorie* (Springer, Berlin), p. 111, footnote1; transl., W. Pauli, *General Principles of Quantum Mechanics,* Springer, Berlin (1980), p. 30, footnote 9.

Pihl, M., 1939, *Der Physiker L. V. Lorenz: Eine Kritische Untersuchung* (Munksgaard, Copenhagen).

Pihl, M., 1972, "The scientific achievements of L.V. Lorenz," Centaurus **17**, 83-94.

Prokhorov, L. V., 2000, "V. A. Fock - The Density of Some Discoveries," Fiz. Elem. Cha. Atom. Yad. **31**, 47-70 [Phys. Part. Nucl. **31**, 22-33].

Ramond, P., 1981, *Field Theory, A Modern Primer* (Benjamin/Cummings, Reading, Mass.), Chapters VII and VIII.

Reiff, R., and A. Sommerfeld, 1902, "Standpunkt der Fernwirkung. Die Elementargesetze," Encykl. Math. Wissen., Band V:2, V. 12, 1-62.

Riemann, B., 1867, "Ein Beitrag zur Elektrodynamik," Ann. der Physik und Chemie **131**, 237-243); "A Contribution to Electrodynamics," Phil. Mag. ser. 4, **34**, 368-372.

Rosenfeld, L., 1957, "The velocity of Light and the Evolution of Electrodynamics," Nuovo Cim. Suppl. ser. 10, **4**, 1630-1669.

Schott, G. A., 1912, *Electromagnetic Radiation* (Cambridge University Press).

Schrödinger, E., 1923, "Über eine bemerkenswerte Eigenschaft der Quantenbahnen eines einzelnen Elektrons," Zeit. für Physik **12**, 13-23.

Schrödinger, E., 1926a, "Quantisierung als Eigenwertproblem I," Ann. der Physik **79**, 361-376. [subm. 27 January 1926, publ. 13 March 1926]

Schrödinger, E., 1926b, "Quantisierung als Eigenwertproblem IV," Ann. der Physik **81**, 109-139. [subm. 21 June 1926, publ. 8 September 1926]

Schrödinger, E., 1978, "Collected Papers on Wave Mechanics," transl. from 2nd German ed., 1928 (Chelsea Publishing, N.Y.).

Schwarzschild, K., 1903, "Zur Elektrodynamik I. Zwei Formen des Princips der kleinsten Action in der Elektronentheorie," Gott. Nach., Math.-phys. Kl., 126-131.

Tricker, R. A. R., 1965, *Early Electrodynamics, the first law of circulation* (Pergamon Press, Oxford)

Van Bladel, J., 1991, "Lorenz or Lorentz?," IEEE Antennas and Propagation Magazine **33**, No. 2, 69.

van der Waerden, B. L., 1973, "From Matrix Mechanics and Wave Mechanics to Unified Quantum Mechanics," in *The Physicist's Conception of Nature*, ed., J. Mehra (Reidel, Dordrecht-Holland), p. 276-291.

Voloshin, M. B., and L. B. Okun, 1978, "Conservation of electric charge," Pis'ma Zh. Eksp. Teor. Fiz. **28**, 156-160. [JETP Lett. **28**, 145-149]

Weber, W., 1878, *Elektrodynamische Maassbestimmungen* (Weidmann, Leipzig)

[A collection of 7 papers, with the same main title, but differing sub-titles, dating from 1846 to 1878. The first part, p. 1-170, is from Abhandlungen der Königlichen Sächsischen Gesellschaft der Wissenschaften, Leipzig (1846)]

Weber, W., 1848, "I. Elektrodynamische Maassbestimmungen," Annalen der Physik und Chemie **73**, 193-240 [shortened version of the 1846 paper published in the Abhandlungen der Königlichen Sächsischen Gesellschaft der Wissenschaften, Leipzig]

Weyl, H., 1919, "Eine neue Erweiterung der Relativitätstheorie," Ann. der Physik **59,** 101-133.

Weyl, H., 1928, *Gruppentheorie und Quantenmechanik* (S. Hirzel, Leipzig), p.87- 88; *Theory of Groups and Quantum Mechanics*, 2nd ed. (1931), transl.H. P. Robertson, Dover, N.Y. (1950), p.100-101.

Weyl, H., 1929a, "Gravitation and the electron," Proc. Nat. Acad. Sci. **15**, 323-334. [subm. 7 March 1929; publ. 15 April 1929] This paper is an early version in English of Weyl (1929b).

Weyl, H., 1929b, "Elektron und Gravitation," Zeit. für Physik **56**, 330-352. [subm. 8 May 1929; publ. 19 July 1929]

Whittaker, E. T., 1951, *History of the Theories of Ether and Electricity, The Classical Theories* (Philosophical Library, New York).

Wiechert, E., 1900, "Elektrodynamische Elementargesetze," Arch. Néerl. Scs., ser. 2, **5**, 549-573.

Yang, C. N., and R. L. Mills, 1954, "Conservation of isotopic spin and isospin gauge invariance," Phys. Rev. **96**, 191-195.

Yang, C. N., 1986, "Hermann Weyl's Contributions to Physics," in *Hermann Weyl, 1885-1985,* ed. K. Chandrasekharan (Springer, Berlin), p. 7-21.

Yang, C. N., 1987, "Square root of minus one, complex phases, and Erwin Schrödinger," in *Schrödinger : centenary celebration of a polymath*, ed. C.W. Kilmister (Cambridge University Press, London), p. 53-64.

Zinn-Justin, J., 1993, *Quantum Field Theory and Critical Phenomena* (Oxford University Press), Chapter 18.

Zumino, B., 1960, "Gauge Properties of Propagators in Quantum Electrodynamics," J. Math. Phys. **1**, 1-7.

**Figure Caption**

Figure 1. Two closed current carrying circuits $C$ and $C'$ with currents $I$ and $I'$, respectively

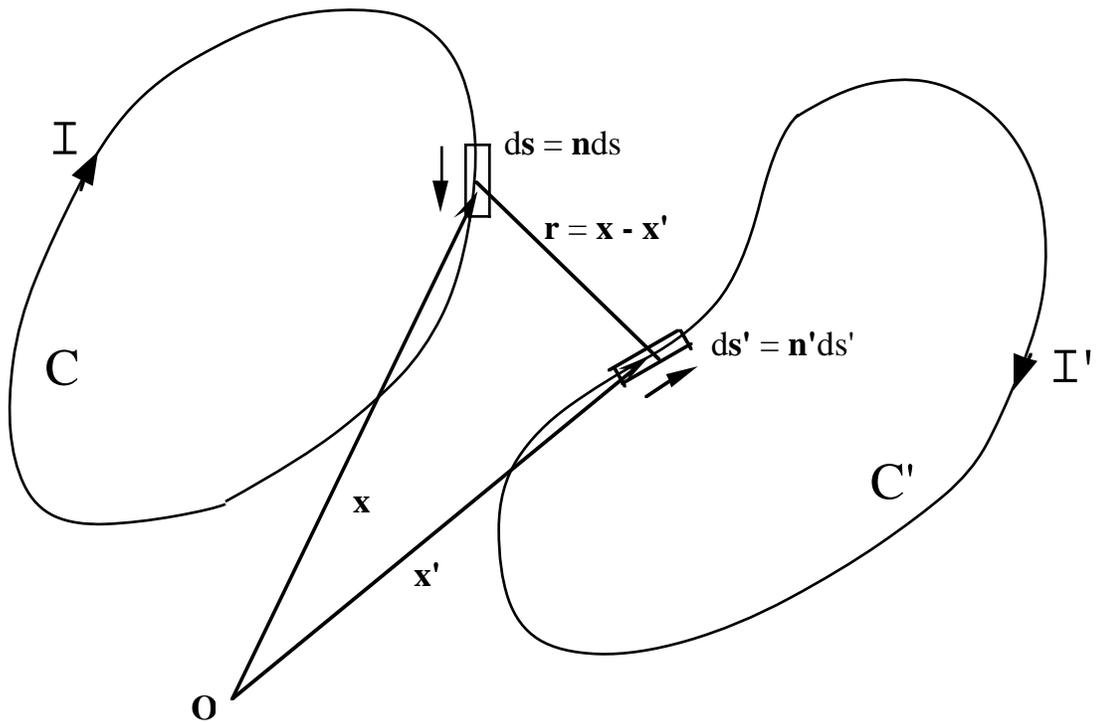

FIGURE 1, JACKSON & OKUN